\documentclass[prc,aps,amssymb,showpacs]{revtex4}
\usepackage{amsmath}
\usepackage{graphicx}
\usepackage[usenames,dvips]{color}
\newcommand{\AP}[3]{Ann.\ Phys.\ {\bf #1},\ #2 (#3)}
\newcommand{\APJ}[3]{AstroPhys.\ J. {\bf #1},\ #2 (#3)}
\newcommand{\CJP}[3]{Canad.\ J.\ Phys. {\bf #1},\ #2 (#3)}

\newcommand{\IJMPE}[3]{Int.\ J.\ Mod.\ Phys.\ {\bf E#1},\ #2 (#3)}

\newcommand{\NP}[3]{Nucl.\ Phys.\ {\bf #1},\ #2 (#3)}
\newcommand{\NPA}[3]{Nucl.\ Phys.\ {\bf A#1},\ #2 (#3)}
\newcommand{\NPB}[3]{Nucl.\ Phys.\ {\bf B#1},\ #2 (#3)}

\newcommand{\RMP}[3]{Rev.\ Mod.\ Phys.\ {\bf #1},\ #2 (#3)}

\newcommand{\SPJETP}[3]{Sov.\ Phys.\ JETP\ {\bf #1},\ #2 (#3)}
\newcommand{\PAN}[3]{Phys.\ Atom.\ Nucl.\ {\bf #1},\ #2 (#3)}

\newcommand{\PLB}[3]{Phys.\ Lett.\ {\bf B#1},\ #2 (#3)}
\newcommand{\PR}[3]{Phys.\ Rep.\ {\bf #1},\ #2 (#3)}
\newcommand{\PRL}[3]{Phys.\ Rev.\ Lett.\ {\bf #1},\ #2 (#3)}
\newcommand{\PRR}[3]{Phys.\ Rev.\ {\bf #1},\ #2 (#3)}

\newcommand{\PRB}[3]{Phys.\ Rev.\ {\bf B#1},\ #2 (#3)}
\newcommand{\PRC}[3]{Phys.\ Rev.\ {\bf C#1},\ #2 (#3)}
\newcommand{\PRD}[3]{Phys.\ Rev.\ {\bf D#1},\ #2 (#3)}

\newcommand{\PPNP}[3]
{Prog.\ Part.\ Nucl.\ Phys.\ {\textbf #1},\ #2 (#3)}
\newcommand{\PTP}[3]{Prog.\ Theor.\ Phys.\ {\bf #1},\ #2 (#3)}

\begin{document}

\title{Neutrino Emission From Direct Urca Processes
in Pion Condensed Quark Matter}
\author{Xuguang Huang$^1$}
\email{huangxg03@mails.tsinghua.edu.cn}
\author{Qun Wang$^2$}
\email{qunwang@ustc.edu.cn}
\author{Pengfei Zhuang$^1$}
\email{zhuangpf@mail.tsinghua.edu.cn}
\affiliation{$^1$Physics Department, Tsinghua University,
Beijing 100084, China\\
$^2$Department of Modern Physics, University of Science and
Technology of China, Hefei, Anhui 230026, China}

\date{\today}

\begin{abstract}
We study neutrino emission from direct Urca processes in pion
condensed quark matter. In compact stars with high baryon density,
the emission is dominated by the gapless modes of the pion
condensation which leads to an enhanced emissivity. While for
massless quarks the enhancement is not remarkable, the emissivity is
significantly larger and the cooling of the condensed matter is
considerably faster than that in normal quark matter when the
mass difference between $u$- and $d$-quarks is sizable.
\end{abstract}
\pacs{12.38.Mh,24.85.+p}
\maketitle

\section {Introduction}
\label{introduction}
Compact stars are born at temperatures $T\gtrsim10^{11}\,\rm{K}$
(about tens of MeV) and rapidly cool down to $10^9\,\rm{K}$ (about
a few MeV) within minutes, and then the cooling process is
dominated by neutrino emission for about $10^5$ years during which
the temperature drops down to $10^6\,\rm{K}$ (about a few KeV)
~\cite{pethick,prakash}. Since the typical density in a compact
star could be several times the normal nuclear density, matter
with strong interaction at such high density and low temperature
must be highly degenerate and may be in a
superfluidity/superconductivity or hybrid
state~\cite{barrois,migdal,migdal1,kaplan}. Neutrino cooling
processes in these phases may behave differently from that in the
normal phase, and the observations of cooling rate can provide
constraints on the phase structure of dense matter.

Many cooling mechanisms have been investigated in literatures. If
a compact star is in the state of nuclear matter, the cooling is
governed by the modified Urca processes and the temperature
dependence of the neutrino emissivity is $\epsilon\sim
T^8$~\cite{chiu,bahcall,tsuruta,friman}. When the proton abundance
is large enough ($\gtrsim 11\%$), the direct Urca processes can
take place and lead to a larger neutrino emissivity $\epsilon\sim
T^6$~\cite{boguta,lattimer}. When pion or kaon condensation is
present, such a fast cooling can be realized as
well~\cite{bahcall,maxwell,muto,brown}. In the nuclear superfluid,
the quasi-particle spectrum of nucleons is gapped and the cooling
mechanism is dramatically changed. At low temperatures $T \ll T_c$
with $T_c$ being the critical temperature of nuclear superfluid,
the Cooper pairs can rarely be broken by thermal fluctuations, the
neutrino emissivity is suppressed by a Boltzman factor
$e^{-\Delta/T}$ with $\Delta$ being the energy gap~\cite{maxwell1,
sedrakian}. When the temperature approaches $T_c$, the effect of
pair breaking and recombination results in an emissivity of
$\epsilon\sim T^7$ for neutrino pair radiation~\cite{flowers}.

Compact stars can be in the state of quark
matter~\cite{itoh,alford0}, when the baryon density is high enough.
The neutrino processes in normal quark matter are similar to that in
nuclear matter. For example, the direct Urca processes including the
$\beta$-decay ($d\rightarrow u+e^-+\bar{\nu}_e$) and the electron
capture ($u+e^-\rightarrow d+\nu_e$) in normal quark matter lead to
an emissivity of $\epsilon\sim T^6$~\cite{iwamoto,burrows0,schafer},
while the modified Urca processes give $\epsilon\sim
T^8$~\cite{burrows}. Quark matter at asymptotically high density is
in the color-flavor locked (CFL) phase~\cite{alford} of color
superconductivity (CSC). The CSC phases at moderate density are
still unclear due to the complicated non-perturbative effect.
Various candidates have been proposed, such as two flavor CSC
(2SC)~\cite{huang}, gapless 2SC~\cite{shovkovy}, gapless CFL
(gCFL)~\cite{alford1}, crystalline CSC~\cite{alford2}, and spin-1
CSC~\cite{iwasaki,schafer1,schmitt}. Neutrino emission from CSC
quark matter is exponentially suppressed at low temperature, if the
quasi-particle spectra are fully
gapped~\cite{alford3,jaikumar,schmitt2,anglani,wang,carter}. Similar
to nuclear matter, the pair breaking and recombination effect
results in an emissivity of $\epsilon\sim T^7$ at temperature close
to $T_c$~\cite{jaikumar2}. The emissivity behaves as $\epsilon\sim
T^{5.5}$ for gCFL~\cite{alford3} and $\epsilon\sim T^{6}$ for g2SC,
see, e.g. Table 1 of Ref.~\cite{schafer}.

At moderate baryon density, the state with spontaneous isospin
symmetry breaking is a possible ground state of quark matter, if the
isospin chemical potential (or equivalently electron chemical
potential) is larger than the pion mass~\cite{son2,he}. In this
state, the nonzero quark-antiquark condensate $\langle\bar d
i\gamma_5 u\rangle$ or $\langle \bar u i\gamma_5 d\rangle$ can be
considered as pion condensate when the coupling between the quark
and antiquark is strong enough. When the temperature of the system
increases, the condensate will melt and there will be, respectively,
BCS and BEC phase transitions in weak and strong coupling regions.
Without making confusion, we call in the following the
quark-antiquark condensate of light flavors as pion condensate. We
emphasize that the pion condensation in the current paper is very
different from the Goldstone boson one in the CFL
phase~\cite{schafer2,buballa} where the Goldstone bosons are
diquark-anti-diquark excitations. In particular the color symmetry
is spontaneously broken in the CFL phase while it is kept in our
case. For example, the kaons $K^0$ and $K^+$ with quantum numbers
$d\bar{s}$ and $u\bar{s}$ in the CFL phase are actually the lowest
excitations and made of $(\bar{u}\bar{s})(ud)$ and
$(\bar{d}\bar{s})(ud)$ respectively, where all diquarks
(anti-diquarks) are in anti-triplet (triplet) in color and flavor
space. In our case the pions are really made of $u\bar{d}$,
$d\bar{u}$ and $u\bar{u}-d\bar{d}$, the same as in vacuum. For
neutrino emission, the role of pion condensation in quark matter is
very different from that in nuclear matter. In nuclear matter, the
pion degrees of freedom open effectively a new channel for nucleon
interaction and result in an enhanced neutrino
emissivity~\cite{bahcall,maxwell}. In quark matter, however, the
pion condensate will suppress, on one hand, the neutrino emission of
pairing quarks due to the cost of breaking the pairs via thermal
excitations, and open, on the other hand, more phase space for
direct Urca processes of those unpaired quarks. The competition
between the two effects determines whether the neutrino emissivity
is suppressed or enhanced.

At moderate baryon density, the system is controlled by light
quarks and the strange quark excitation is not yet important. In
this paper, we will calculate the neutrino emissivity and cooling
rate through direct Urca processes in pion condensed quark matter
with only two flavors. Since the typical baryon chemical potential
$\mu_B$ is of the order of $1$ GeV and the isospin chemical
potential $\mu_I$ is about $100-250$ MeV in compact stars, the
mismatch between the two Fermi surfaces of the pairing $u(\bar u)$
and $\bar d(d)$ quarks is large enough and the pion superfluid
will be in gapless state when the condensate $\Delta$ satisfy the
condition $\mu_B/3>\Delta$. We will focus on the gapless pion
condensation.

The paper is organized as follows. In Section \ref{emissivity} we
derive the general formulation for the emissivity in pion
condensed quark matter. The calculations of the emissivity and
cooling rate for positive energy excitations which are dominant at
low temperature are presented in Sections \ref{p-em} and
\ref{cooling_rate}. The contribution from negative energy
excitations to the emissivity is discussed in Section \ref{n-em}.
The summary is given in the final section.
Our units are $\hbar=k_B=c=1$ except particular specification.
As a convention, we denote a 4-momentum as $K^\mu =(k_0,\mathbf{k})$,
and its 3-momentum magnitude as $k=|\mathbf{k}|$.

\section {General formulation for neutrino emissivity}
\label{emissivity}
At a typical temperature for a relatively aging neutron star, the
neutrino mean free path is larger than the star radius, and there
is no accumulation of neutrinos inside the star. In this case, the
chemical potential for neutrinos can be taken to be zero. We
assume $\beta$-equilibrium, so the chemical potentials satisfy
$\mu_d=\mu_u+\mu_e$ for the quark chemical potentials $\mu_u$ and
$\mu_d$ and the electron chemical potential $\mu_e$. Since the
pion condensation favors a high isospin or equivalently electron
chemical potential, we assume that $\mu_e$ is of the same order of
$\mu_u$ and $\mu_d$.
\begin{figure}[!htb]
\includegraphics[width=7cm]{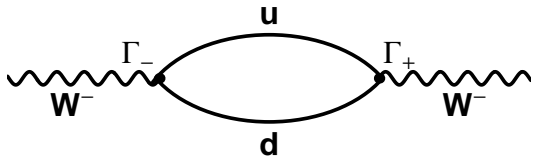}
\caption{The one-loop diagram of W-boson polarization tensor.}
\label{w-polar}
\end{figure}

It is convenient to use the Fermi effective model of weak
interactions to describe the Urca processes where the
characteristic energy scale is about a few hundred MeV and much
less than the mass of W-bosons. The interaction Lagrangian of the model is
\begin{equation}
\mathcal{L}_{\rm int} = \frac{G}{\sqrt{2}}J^{\mu}J_{\mu}^{\dagger}
\end{equation}
with the weak current
\begin{equation}
J^{\mu}(x) = \bar{\nu}\gamma^{\mu}(1-\gamma_{5})e +
\bar{u}\gamma^{\mu}(1-\gamma_{5})d ,
\label{eq:jmu}
\end{equation}
where $u,d,e,\nu$ denote the fields for $u$ and $d$ quarks,
electrons and neutrinos, respectively, and
$G=G_F\cos\theta_C\approx1.13488\times10^{-11}$ ${\rm MeV}^{-2}$
is the four-fermion coupling constant with the Cabibbo angle $\theta_C$.

The neutrino emissivity, defined as the total energy per unit time
and per unit volume carried away by neutrinos and anti-neutrinos,
can be written as
\begin{subequations}
\label{emss}
\begin{eqnarray}
\epsilon _{\nu} & = & G^2 \int\frac{d^3{\bf p}_e}{(2\pi)^32E_{e}}
\int\frac{d^3{\bf p}_\nu}{(2\pi)^3 2E_{\nu}} E_{\nu}
n_B(-E_{e}+\mu_e+E_{\nu} ) n_F(E_{e}-\mu_e) \nonumber\\
&& \times L^{\lambda\sigma}(P_\nu,P_e)\mathrm{Im}\Pi^R_{\lambda\sigma}
(E_{e}-\mu_e-E_{\nu},\mathbf{p}_e-\mathbf{p}_\nu),\\
\epsilon _{\bar{\nu}}& = & -G^2 \int\frac{d^3{\bf
p}_e}{(2\pi)^32E_{e}} \int\frac{d^3{\bf p}_\nu}{(2\pi)^3 2E_{\nu}}
E_{\nu}
n_B(E_{e}-\mu_e+E_{\nu} ) n_F(\mu_e-E_{e}) \nonumber\\
&& \times
L^{\lambda\sigma}(P_\nu,P_e)\mathrm{Im}\Pi^R_{\lambda\sigma}
(E_{e}-\mu_e+E_{\nu},\mathbf{p}_e+\mathbf{p}_\nu),
\end{eqnarray}
\end{subequations}
where $P_i=(E_i,{\mathbf p}_i)$ for $i=u,d,e,\nu$ are on-shell
4-momenta for quarks and leptons. The energies are
$E_i \equiv E_{p_i}=\sqrt{p_i^2+m_i^2}$ with the masses $m_i$ for $i$
and with $m_{\nu}=0$ and $m_{e}\approx 0$.
Here $n_B(x)=(e^{x/T}-1)^{-1}$ and $n_F(x)=(e^{x/T}+1)^{-1}$ are the
Bose-Einstein and Fermi-Dirac distribution functions, respectively.
$L^{\lambda\sigma}(P_e,P_\nu)$ denotes the leptonic tensor and
$\Pi^R_{\lambda\sigma}(Q)$ the retarded polarization tensor for
W-bosons. The leptonic tensor reads
\begin{eqnarray}
L^{\lambda\sigma}(P_\nu,P_e)
&=&\mathrm{Tr}[\gamma^\lambda(1-\gamma_5)\gamma\cdot
P_e\gamma^\sigma(1-\gamma_5)\gamma\cdot P_\nu]\nonumber\\
&=& 8 [P_e^\lambda P_\nu^\sigma+P_\nu^\lambda P_e^\sigma-P_e\cdot
P_\nu g^{\lambda\sigma}+i\epsilon^{\lambda\alpha\sigma\beta}
P_{e\alpha}P_{\nu\beta}].
\end{eqnarray}
As illustrated in Fig. \ref{w-polar} the $W$-boson polarization
tensor can be written as
\begin{equation}
\label{pi1} \Pi^{\lambda\sigma}(q_0,\mathbf{q}) = T\sum
_n\int\frac{d^3\mathbf{p}_u}{(2\pi )^3}\mathrm{Tr} [\Gamma^\lambda_-
S(p_{u0},\mathbf{p}_u)\Gamma^\sigma_+ S(p_{d0},\mathbf{p}_d)],
\end{equation}
where $p_{d0}= p_{u0}+q_0$, $\mathbf{p}_d=\mathbf{p}_u+\mathbf{q}$, and
$\Gamma_\pm^\lambda=\gamma^\lambda(1-\gamma_5)\tau_\pm$ with
the ladder operators in flavor space $\tau_\pm=(\tau_1\pm i\tau_2)/2$.
The bosonic and fermionic Matsubara frequencies are given by
$q_0=i 2l\pi T$ and $p_{u0}=i(2n+1)\pi T$ ($l,n$ are integers) respectively.
The retarded polarization tensor $\Pi^{R}$ is obtained from the above
after taking the analytic extension for the Matsubara frequency
$q_0=i 2l\pi T\rightarrow q_0+i0^+$, where the second $q_0$ is real.
We can then extract the imaginary part Im$\Pi^{R}$ from $\Pi^{R}$.
The full propagator for quarks
\begin{eqnarray}
\label{propagator}
S(K)=\left[\begin{array}{cc} S_{uu}(K) & S_{ud}(K) \\
S_{du}(K) & S_{dd}(K)
\end{array}\right]
\end{eqnarray}
in flavor space can be derived from its inverse~\cite{he},
\begin{equation}
\label{s-inverse}
S^{-1}(K)=\left(\begin{array}{cc}\gamma^\nu K_\nu+\mu_u\gamma_0-m &
i\gamma_5\Delta^- \\ i\gamma_5\Delta^+ & \gamma^\nu
K_\nu+\mu_d\gamma_0-m
\end{array}\right),
\end{equation}
where $m$ is the average effective quark mass, and
$\Delta^+\sim\langle\bar{d}i\gamma_5u\rangle$ and
$\Delta^-\sim\langle\bar{u}i\gamma_5d\rangle$ are the charged pion
condensates. Without loss of generality, we assume real condensates
and let $\Delta^+=\Delta^-=\Delta$. In this case, the four matrix
elements of $S$ are explicitly expressed as
\begin{eqnarray}
S_{uu} & = & \sum _{a,r=\pm} B_{r}^{a}(k)\frac{\Lambda_{k}^{a}\gamma_{0}}{k_{0}+\mu_{B}/3+rE_{k}^{a}},\\
S_{dd} & = & \sum _{a,r=\pm} B_{-r}^{a}(k)\frac{\Lambda_{k}^{-a}\gamma_{0}}{k_{0}+\mu_{B}/3+rE_{k}^{a}},\\
S_{ud} & = & -i \sum _{a,r=\pm} r\sqrt{B_{+}^{a}(k)B_{-}^{a}(k)}\frac{\Lambda_{k}^{a}\gamma_{5}}{k_{0}+\mu_{B}/3+rE_{k}^{a}},\\
S_{du} & = & -i \sum _{a,r=\pm} r\sqrt{B_{+}^{a}(k)B_{-}^{a}(k)}\frac{\Lambda_{k}^{-a}\gamma_{5}}{k_{0}+\mu_{B}/3+rE_{k}^{a}},
\end{eqnarray}
where the summation over $a$ (and $b$ in the following) is
for positive and negative energy excitations for quarks,
and the that over $r$ (and $s$ in the following) is for quarks and
anti-quarks. The Bogoliubov coefficients $B_{r}^{a}(k)$,
energy projectors $\Lambda ^a_k$ and energies $E_k^a$ are defined by
\begin{eqnarray}
B_{r}^{a}(k) & = &
{1\over2}\left[1-ar\frac{E_{k}+a\delta\mu}{E_{k}^{a}}\right],\nonumber \\
\Lambda ^a_k & = & {1\over2}\left[1+a\frac{\gamma_0({\gamma\cdot\bf
k}+m)}{E_k}\right], \nonumber \\
E_k^a & = & \sqrt{(E_k+a\delta\mu)^2+\Delta^2}.
\end{eqnarray}
We use $\delta\mu \equiv (\mu_d - \mu_u )/2 = -\mu _I/2=\mu_e/2$ and
$\mu \equiv \mu_B/3$ to replace $\mu_B$ and $\mu_e$,
where $\mu_{I,B,e}$ are chemical potentials for
the isospin, baryons and electrons respectively.
From the poles of the quark propagators,
the quasi-quark energy is then given by $E_k^a+r\mu$.
The chemical potentials for $u$ and $d$ quarks can be expressed in terms of
$\delta\mu$ and $\mu$ as $\mu _u=\mu -\delta\mu$ and $\mu_d=\mu +\delta\mu$.

Substituting the quark propagator (\ref{propagator}) into the
W-boson polarization tensor (\ref{pi1}) and performing the trace
in color and flavor space, we obtain
\begin{eqnarray}
\label{pi}
\Pi^{\lambda\sigma}(q_0,\mathbf{q}) & = &
N_c T \sum _n\int\frac{d^3\mathbf{p}_u}{(2\pi )^3}\mathrm{Tr}
[\gamma^\lambda (1-\gamma_5) S_{uu}(p_{u0},\mathbf{p}_u)\gamma^\sigma
(1-\gamma_5) S_{dd}(p_{d0},\mathbf{p}_d)] \nonumber\\
&=& N_c \sum _{a,b,r,s=\pm} \int\frac{d^3\mathbf{p}_u}{(2\pi )^3 4E_{u}E_{d}} H_{ab}^{\lambda\sigma}(P_u,P_d)
B_{r}^{a}(p_u) B_{-s}^{-b}(p_d)
\frac{n_F(-rE_{u}^{a}-\mu )-n_F(-sE_{d}^{-b}-\mu )}
{q_0-rE_{u}^{a}+sE_{d}^{-b}} ,
\end{eqnarray}
with the quark tensor $H_{ab}^{\lambda\sigma}(P_u,P_d)$ given by
\begin{eqnarray}
H_{ab}^{\lambda\sigma}(P_u,P_d) & = & 4E_{u}E_{d} \mathrm{Tr}
[\gamma^\lambda(1-\gamma_5)\Lambda^a_{p_u}\gamma_0
\gamma^\sigma(1-\gamma_5)\Lambda^{b}_{p_d}\gamma_0]\nonumber\\
&=& 8 [P_{au}^\lambda P_{bd}^\sigma+P_{bd}^\lambda
P_{au}^\sigma-P_{au} \cdot P_{bd}
g^{\lambda\sigma}+i\epsilon^{\lambda\ \sigma}_{\ \alpha\
\beta}P_{au}^{\alpha}P_{bd}^{\beta}],
\end{eqnarray}
where $P_{au}=(E_{u},a\mathbf{p}_{u})$ with $a=+,-$ and
$P_{bd}=(E_{d},b\mathbf{p}_{d})$ with $b=+,-$ for positive and
negative energy excitations respectively. We have also used the
number of colors $N_c$ which is finally set to 3 in our calculation.
Using $\mathrm{Im}\Pi^{\lambda\sigma}_R(q_0,{\bf
q})=\mathrm{Im}\Pi^{\lambda\sigma}(q_0+i0^+,{\bf q})$, the imaginary
part of the retarded polarization tensor of W-bosons can be read out
directly,
\begin{eqnarray}
\label{impi}
\mathrm{Im}\Pi^{\lambda\sigma}_R(q_0,{\bf q}) & = &
\pi N_c \sum _{a,b,r,s=\pm} \int\frac{d^3\mathbf{p}_u}{(2\pi )^3 4E_{u}E_{d}}
H_{ab}^{\lambda\sigma}(P_u,P_d)
B_{r}^{a}(p_u) B_{-s}^{-b}(p_d)n_B^{-1}(-q_0)\nonumber\\
&&\times n_F(-rE_{u}^{a}-\mu )n_F(sE_{d}^{-b}+\mu )
\delta (q_0-rE_{u}^{a}+sE_{d}^{-b}) \nonumber\\
& = & -\pi N_c \sum _{a,b,r,s=\pm} \int\frac{d^3\mathbf{p}_u}{(2\pi )^3 4E_{u}E_{d}} H_{ab}^{\lambda\sigma}(P_u,P_d)
B_{r}^{a}(p_u) B_{-s}^{-b}(p_d)n_B^{-1}(q_0)\nonumber\\
&&\times n_F(rE_{u}^{a}+\mu )n_F(-sE_{d}^{-b}-\mu )
\delta (q_0-rE_{u}^{a}+sE_{d}^{-b}),
\end{eqnarray}
where the first equality is for the neutrino emissivity with
$q_0=E_{e}-\mu_e-E_{\nu}$ and the second for the anti-neutrino one
with $q_0=E_{e}-\mu_e+E_{\nu}$. Here we
have used the identity
$n_F(x)-n_F(y)=n_F(-x)n_F(y)/n_B(-x+y)=-n_F(x)n_F(-y)/n_B(x-y)$.

Substituting the imaginary polarization (\ref{impi}) into the
neutrino emissivity (\ref{emss}), we arrive at
\begin{eqnarray}
\epsilon _{\nu} & = & 2N_c\sum _{a,b,r,s=\pm}\int\frac{d^3\mathbf{p}_e}{(2\pi)^3 2E_e}
\int \frac{d^3\mathbf{p}_\nu}{(2\pi)^3 2E_\nu}
\int\frac{d^3\mathbf{p}_u}{(2\pi)^32 E_u}
\int\frac{d^3\mathbf{p}_d}{(2\pi)^3 2E_d}
E_\nu (2\pi)^4 \delta(E_{e}-\mu_e-E_{\nu}-rE_u^a + sE_d^{-b})
\nonumber\\
&& \times \delta^3(\mathbf{p}_e-\mathbf{p}_\nu+\mathbf{p}_u-\mathbf{p}_d)
B_{r}^{a}(p_u) B_{-s}^{-b}(p_d) n_F(E_e-\mu_e) n_F(-rE_{u}^{a}-\mu )
n_F(sE_{d}^{-b}+\mu ) |M_{ab}|^2 ,\nonumber \\
\epsilon _{\bar{\nu}} & = & 2N_c\sum _{a,b,r,s=\pm}\int\frac{d^3\mathbf{p}_e}{(2\pi)^3 2E_e}
\int \frac{d^3\mathbf{p}_\nu}{(2\pi)^3 2E_\nu}
\int\frac{d^3\mathbf{p}_u}{(2\pi)^32 E_u}
\int\frac{d^3\mathbf{p}_d}{(2\pi)^3 2E_d}
E_\nu (2\pi)^4 \delta(E_{e}-\mu_e+E_{\nu}-rE_u^a + sE_d^{-b}) \nonumber\\
&& \times \delta^3(\mathbf{p}_e+\mathbf{p}_\nu+\mathbf{p}_u-\mathbf{p}_d)
B_{r}^{a}(p_u) B_{-s}^{-b}(p_d) n_F(\mu_e-E_e) n_F(rE_{u}^{a}+\mu )
n_F(-sE_{d}^{-b}-\mu ) |M_{ab}|^2 ,
\label{e2ab}
\end{eqnarray}
where the shorthand notation $|M_{ab}|^2$ is the spin-averaged
matrix element for the $\beta$-decay or the electron capture
processes~\cite{iwamoto},
\begin{equation}
\label{amplitude} |M_{ab}|^2 = \frac{G^2}{4}
L_{\lambda\sigma}(P_e,P_\nu)H_{ab}^{\lambda\sigma}(P_u,P_d) =64
G^2(P_e\cdot P_{au})(P_\nu\cdot P_{bd}).
\end{equation}
Here we have suppressed energy projection indices $a,b$ in quark
momenta as before. When $a=+$ and $b=+$ for $u$ and $d$ quarks
respectively, $|M_{++}|^2$ is just the matrix element of the Urca
processes in vacuum. For other combinations of $a$ and $b$, it
represents the matrix elements with at least one negative energy
excitation for $u,d$ quasi-quarks.

\section{Neutrino emission with positive energy excitations}
\label{p-em}
It is well known that the negative energy excitations are normally
neglected at low temperature due to their strong suppression
relative to the positive excitations, see, e.g. Eqs. (21-24) for
fully gapped phases in Ref.~\cite{schmitt2}. This can be
understood from the analysis of the parts involving Bogoliubov
coefficients and quark number distributions in Eq. (\ref{e2ab}),
\begin{equation}
B_{r}^{a}(p_u) B_{-s}^{-b}(p_d)n_F(\mp rE_{u}^{a}\mp\mu )n_F(\pm sE_{d}^{-b}\pm\mu ),
\end{equation}
where the upper and lower signs correspond to neutrinos and
anti-neutrinos respectively. Let us consider the part related to
the negative energy excitations for $u$ quarks,
$B_r^-(p_u)n_F(-rE^-_u-\mu)$ and $B_r^-(p_u)n_F(rE^-_u+\mu)$. Note
that the dominant contribution to the emissivity comes from the
phase space near gapless nodes which correspond to the vanishing
arguments of quark distribution functions by selecting the branch
with $r=-$, see Appendix \ref{asymp}. The gapless node for $u$
quarks in this case is located at
$E_u^0=\sqrt{\mu^2-\Delta^2}+\delta\mu$, then we obtain
$B_-^-(p_u) = \left(1-\sqrt{\mu^2-\Delta^2}/\mu\right)/2$ which
approaches zero for large $\mu$ or small $\Delta$. The same
conclusion is also true for $d$ quarks. The above arguments are
valid only at low temperatures $T\ll\mu_{u,d}$ where the dominant
contribution of phase space integrals comes from the region near
gapless nodes.

In this section we calculate the neutrino emissivity by dropping
out the negative energy excitations. We will briefly discuss the
contribution from the negative energy excitations in Section \ref{n-em}.

\begin{figure}[!htb]
\includegraphics[width=7cm]{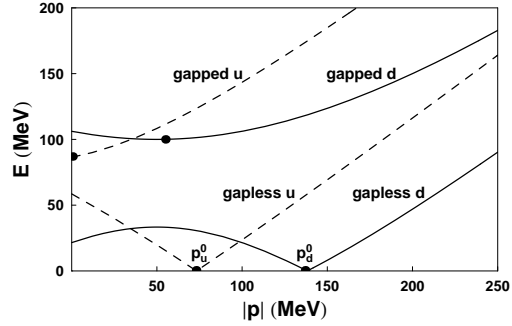}
\caption{The schematic dispersion relations for gapped and gapless
quasi-quarks.} \label{dispersion1}
\end{figure}

When quarks are treated as free and massless constituents in
normal quark matter, phase space for direct Urca processes
vanishes, since quarks and electrons are collinear in momenta on
their Fermi surfaces~\cite{iwamoto}. To get non-vanishing phase
space which requires that the Fermi momenta of $u$ and $d$ quarks
and electrons should satisfy a triangular relation, one must take
into account nonzero quark masses and/or modified quark dispersion
relations due to interactions and/or pairings between
(anti-)quarks. We will consider these effects together. When the
baryon chemical potential is high enough, the dispersion relations
of quasi-quarks are gapless, the neutrino emissivity is dominated
by the quasi-quarks around the gapless momenta. For $u$ and $d$
quarks, the gapless momenta are $p_u^0 =
(1-\kappa)\sqrt{(E_u^0)^2-m_u^2}$ and $p_d^0 =
(1-\kappa)\sqrt{(E_d^0)^2-m_d^2}$ with the gapless energies
$E_u^0=\sqrt{\mu^2-\Delta^2}-\delta\mu$ and
$E_d^0=\sqrt{\mu^2-\Delta^2}+\delta\mu$. The dispersion relations
of quasi-quarks are schematically shown in Fig.\
\ref{dispersion1}. We have included the interaction among quarks
by introducing a phenomenological constant $\kappa$ which reflects
the reduction of the Fermi momenta and can be derived in Landau's
Fermi liquid theory or in field theory. In perturbative QCD we
have $\kappa = 2\alpha_s/(3\pi)$ with $\alpha_s$ being the strong
coupling constant~\cite{baym}, and in the NJL model it can be
expressed as $\kappa = 4g\mu_B^2/(3\pi^2)$ with the four-fermion
coupling constant $g$~\cite{wang}. In the normal phase with
vanishing pion condensate, $E_{u,d}^0$ and $p_{u,d}^0$ are just
the Fermi energies and momenta, i.e. $E_{u,d}^0=\mu_{u,d}$ and
$p_{u,d}^0=(1-\kappa)\mu_{u,d}$.

At the gapless momenta, the angles between $u$-quark and electron
momenta and between $u$- and $d$-quark momenta can be written as
\begin{eqnarray}
\cos\theta_{ue}^0 & = & \frac{(p_d^0)^2-(p_u^0)^2-\mu_e^2}{2\mu_ep_u^0},\nonumber\\
\cos\theta_{ud}^0 & = & \frac{(p_d^0)^2+(p_u^0)^2-\mu_e^2}{2p_d^0p_u^0}.
\end{eqnarray}
Substituting them into Eq. (\ref{amplitude}), one obtains
approximately
\begin{eqnarray}
|M_{++}|^2 \approx 64 G^2 E_e E_u E_\nu
E_d \left(1-\frac{p_u^0}{E_u^0}\cos{\theta_{ue}^0}\right)
\left(1-\frac{p_d^0}{E_d^0}\cos{\theta_{d\nu}}\right),
\end{eqnarray}
where $\theta_{d\nu}$ is the angle between neutrino and $d$-quark
momenta. In the brackets the $u$- and $d$-quark energies and
momenta and the angle between $u$-quark and electron momenta are
replaced by the corresponding values at the gapless
nodes. Taking into account the relation $\mu_d=\mu_u+\mu_e$, the
$\delta$-functions in Eq. (\ref{e2ab}) can be simplified as
$\mu_e/(p_u p_d)\delta (\cos\theta_{ud}-\cos\theta_{ud}^0)$ to the
leading order of $p_\nu/\mu_e$, since the higher order
contribution is small in the kinetic range of the processes. Using
the above approximation, the neutrino and anti-neutrino emissivity
can be expressed in the following form
\begin{eqnarray}
\label{e3ab}
\epsilon_\nu &\approx& 16\pi N_c G^2 \mu_e E_u^0 E_d^0
\left(1-\frac{p_u^0}{E_u^0}\cos{\theta_{ue}^0}\right) J
\sum_{r,s=\pm} \int_0^\infty dE_\nu E^3_\nu\int_{m}^\infty
dE_u\int_{m}^\infty dE_d\nonumber\\
&&\times B^+_r(p_u)B^-_{-s}(p_d)n_F(E_\nu+rE_u^+ - sE_d^-)n_F(-rE_{u}^{+}-\mu)
n_F(sE_{d}^{-}+\mu),\nonumber\\
\epsilon_{\bar{\nu}} &\approx& 16\pi N_c G^2 \mu_e E_u^0 E_d^0
\left(1-\frac{p_u^0}{E_u^0}\cos{\theta_{ue}^0}\right) J
\sum_{r,s=\pm}\int_0^\infty dE_\nu E^3_\nu \int_{m}^\infty
dE_u\int_{m}^\infty dE_d\nonumber\\
&& \times B^+_r(p_u)B^-_{-s}(p_d)n_F(E_\nu-rE_u^+ + sE_d^-)n_F(rE_{u}^{+}+\mu)
n_F(-sE_{d}^--\mu),
\end{eqnarray}
where we have used $d p_u d p_d =d E_u d E_d E_u E_d/(p_u p_d)$,
and the angular integration $J$ has been evaluated as
\begin{equation}
\label{angle}
J=\int\frac{d\Omega_\nu}{(2\pi)^3}\int\frac{d\Omega_u}{(2\pi)^3}\int\frac{d\Omega_d}{(2\pi)^3}
(1-\frac{p_d^0}{E_d^0}\cos\theta_{d\nu})\delta(\cos\theta_{ud}-\cos\theta_{ud}^0)
=\frac{1}{16\pi^6}.
\end{equation}

To further simplify the emissivity, it is convenient to introduce
the dimensionless variables
\begin{equation}
\label{dimensionless}
x=\frac{E_u+\delta\mu}{T},\,
y=\frac{E_d-\delta\mu}{T},\, u=\frac{\mu}{T},\,
v=\frac{E_\nu}{T},\, w=\frac{\Delta}{T},
\end{equation}
and relax the integration ranges of $x, y$ to $(-\infty,\infty )$,
since the integrations are dominated by the gapless nodes.
Dropping out odd terms of $x$ and $y$ from the integrations, and
noting that the summation over $r$ and $s$ is symmetric,
$\epsilon_\nu$ and $\epsilon_{\bar\nu}$ in (\ref{e3ab}) are
approximately the same. Therefore, we have
\begin{equation}
\label{e4}
\epsilon = \epsilon_\nu+\epsilon_{\bar \nu} \approx
\frac{6}{\pi^5} G^2\mu_e E_u^0
E_d^0\left(1-\frac{p_u^0}{E_u^0}\cos{\theta_{ue}^0} \right) T^6
F(u,w),
\end{equation}
with
\begin{eqnarray}
\label{Frs}
F(u,w) &=&
\sum_{r,s=\pm}F_{rs}(u,w),\nonumber\\
F_{rs}(u,w) &=& \int_0^\infty dx dy dv
v^3\frac{1}{e^{-r\sqrt{x^2+w^2}-u}+1}\frac{1}{e^{s\sqrt{y^2+w^2}+u}+1}
\frac{1}{e^{-s\sqrt{y^2+w^2}+r\sqrt{x^2+w^2}+v}+1}.
\end{eqnarray}
While the two integrals in (\ref{e3ab}) for neutrinos and
anti-neutrinos are different in general case, see Appendix
\ref{exact}, they approach to the same value at low temperature,
because the gapless momenta scaled by temperature are far from
zero providing high effective Fermi surfaces. In this case, the
integrals perfectly center at the gapless momenta and other
contributions quickly damp out. However this is not the case at
$T_c$. The difference and coincidence in the neutrino and
anti-neutrino emissivities also occur in fully gapped
superfluid/supeconducting states.

Eq. (\ref{e4}) is our main result, which presents the neutrino
emissivity $\epsilon$ as a function of temperature $T$, chemical
potentials $\mu_B$ and $\mu_e$, the effective quark mass $m_{u,d}$,
and the pion condensate $\Delta$.

To clearly see the effects of the pion condensate, we consider two
limit cases. One is the high temperature limit with $T$ close to
$T_c$ where the condensate is small, $\Delta\ll\mu_u, \mu_d, \mu_e,
T$, and its temperature dependence can be modeled as
$\Delta=\Delta_0\sqrt{1-(T/T_c)^2}$ with $\Delta_0$ being the pion
condensate at $T=0$. The other is the low temperature limit with
$T\ll T_c$. Since the chiral symmetry is largely restored in the
phase of pion condensation, the effective quark mass is small
compared with the quark chemical potential (actually,
$m/\mu\propto\sqrt{\kappa}$), we can assume $m\ll\mu_u,\mu_d,\mu_e$.

In the limit of high temperature, we can expand $E_{u,d}^0$,
$p_{u,d}^0$ and $\cos\theta^0_{ue}$ in terms of $\kappa$,
$m_{u,d}^2/\mu^2$ and $\Delta ^2/\mu^2$, see Appendix \ref{app1}.
By keeping only the leading order and considering the limit
$\mathrm{Lim}_{w\rightarrow 0} F(u,w)=457\pi^6/5040$~\cite{morel},
we obtain the emissivity near $T_c$,
\begin{equation}
\label{e5}
\epsilon \approx
\left(1-\frac{\Delta^2}{\mu_u\mu_d}\right)\frac{1371}{2520}\pi
G^2\mu_e\mu_u\mu_d\left[\left(1+\frac{\mu_d}{\mu_u}\right)\kappa
+\frac{m_d^2-m_u^2}{2\mu_u\mu_e}\right]T^6.
\end{equation}
Comparing it with the result in normal quark matter with massless
quarks~\cite{iwamoto},
\begin{eqnarray}
\label{e0}
\epsilon_0 & \approx & \frac{1371}{2520}\pi
G^2\mu_e\mu_u\mu_d
\left(1-\frac{p_u^0}{E_u^0}\cos{\theta_{ue}^0} \right) T^6 \nonumber\\
&\approx& \frac{1371}{2520}\pi
G^2\mu_e\mu_u\mu_d\left(1+\frac{\mu_d}{\mu_u}\right)\kappa T^6,
\end{eqnarray}
we conclude that, near $T_c$ a small pion condensate reduces the
emissivity slightly but the correction due to the quark mass
splitting may enhance it significantly.

In the low temperature limit, at $\mu_B=0\ (u=0)$ we can recover
the fully gapped phase and the emissivity has an exponential
suppression factor $e^{-\Delta/T}$ (see, e.g.~\cite{schmitt2}).
For non-vanishing $\mu_B$ or $u$, such a suppression is absent due
to the gapless modes appeared in quark number distribution
functions. In the case with $u > w \gg 1$, we derived in Appendix
\ref{asymp} the asymptotic behavior of $F_{rs}(u,w)$ which can be
summarized as
\begin{eqnarray}
F_{++} & \sim & we^{-u-w},\nonumber\\
F_{+-} & \sim & w^{1/2} (u^2-w^2)^{1/2} e^{-u-w},\nonumber\\
F_{-+} & \sim & w^{1/2} (u^2-w^2)^{9/2} e^{-u-w},\nonumber\\
F_{--} & \sim & O(1).
\label{asymptote}
\end{eqnarray}
It is easy to see that the term $F_{--}(u,w)$ dominates $F(u,w)$
due to the gapless modes in both $u$ and $d$ quark number
distribution functions. Therefore, at low temperature the
emissivity is of the same order as that in normal quark matter.
\begin{figure}[!htb]
\includegraphics[width=7cm]{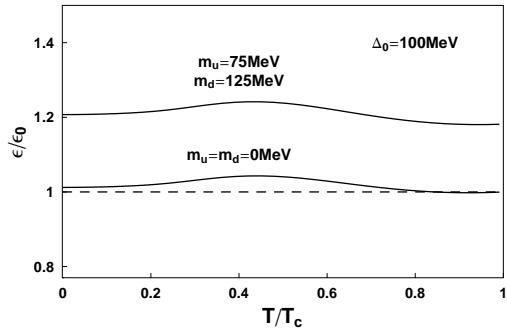}
\caption{The scaled neutrino emissivity $\epsilon/\epsilon_0$ as
functions of the scaled temperature $T/T_c$ for massless and
massive quarks. } \label{e-T}
\end{figure}

To do numerical calculation, we choose the parameters $\mu_B=900$
MeV, $\mu_I=150$ MeV, $\kappa=2/(3\pi)$ and $\Delta_0=100$ MeV.
These are typical values for the possible pion condensed quark
matter in compact stars. The numerical result for the ratio
$\epsilon/\epsilon_0$ as functions of $T/T_c$ is illustrated in
Fig.\ \ref{e-T} where the critical temperature is given by
$T_c=0.57\Delta_0$ according to the BCS theory. When the chiral
symmetry is fully restored, the quarks are massless, and the
difference between pion condensed and normal quark matter is small
at low temperature and becomes considerable at intermediate
temperature. At high temperature $T\rightarrow T_c$, the
emissivity in pion condensed matter is a little lower than that in
normal matter, just as we expected above. However, in general case
with nonzero baryon and isospin chemical potentials, the effective
quark masses $m_u$ and $m_d$ for $u$- and $d$-quarks are not the
same, there should exist a quark mass splitting induced by the
isospin symmetry breaking. In the massive case with $m_u=75$ MeV
and $m_d=125$ MeV, the enhancement factor for the emissivity
becomes 1.2 in the whole temperature range. A similar behavior was
found in a two-flavor gapless color
superconductor~\cite{jaikumar}.

\section{Cooling Rate}
\label{cooling_rate}
\begin{figure}[!htb]
\includegraphics[width=7cm]{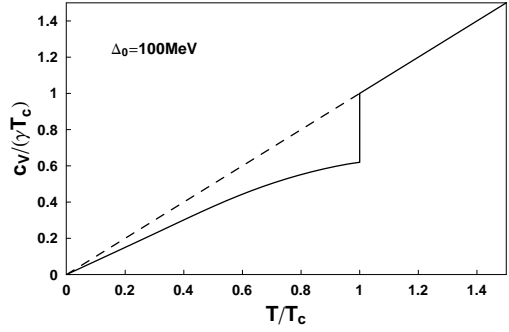}
\caption{The scaled specific heat $c_V/(\gamma T_c)$ as functions
of scaled temperature $T/T_c$. The straight dashed line is the
result in normal quark matter, and the solid line is the result
in the pion condensation. There is a jump at the phase transition
point.}
\label{cvT}
\end{figure}

In this section, we compute the cooling rate due to neutrino
emission. To this end, we integrate the energy equation
$\epsilon(T)=-c_V(T)dT/dt$ and obtain
\begin{equation}
\label{rate}
t-t_0=-\int_{T_0}^TdT'\frac{c_V(T')}{\epsilon(T')},
\end{equation}
where $T_0$ is the initial temperature at time $t_0$. From the
entropy density
\begin{eqnarray}
s&=&-2N_c\sum _{a,r=\pm}\int\frac{d^3{\mathbf k}}{(2\pi)^3}\Big\{n_F(E_k^a+r\mu)
\ln n_F(E_k^a+r\mu)\nonumber\\
&&+[1-n_F(E_k^a+r\mu)]\ln [1-n_F(E_k^a+r\mu)]\Big\},
\end{eqnarray}
the specific heat $c_V$ is given by
\begin{eqnarray}
c_V(T)&=&T\frac{\partial s}{\partial T}\nonumber\\
&=&2N_c\sum _{a,r=\pm}\int\frac{d^3{\bf k}}{(2\pi)^3}n_F(E_k^a+r\mu)n_F(-E_k^a-r\mu)\nonumber\\
&&\times\left[\frac{1}{T^2}(E_k^a+r\mu)^2 -\frac{1}{T}(E_k^a+r\mu)
\frac{\partial E_k^a}{\partial \Delta}\frac{\partial \Delta}{\partial T}\right]
\nonumber\\
&\approx&T\frac{N_c}{\pi^2}\sum _{a=u,d}p_a^0E_a^0\left[
K(u,w)+\frac{\Delta_0^2}{T_c^2}G(u,w)\right],
\label{cv1}
\end{eqnarray}
where we have used the relation
$\Delta(T)=\Delta_0\sqrt{1-(T/T_c)^2}$ and made the
approximation that the integral is dominated by the gapless nodes,
and the functions $K(u,w)$ and $G(u,w)$ are defined by
\begin{eqnarray}
K(u,w)&=&\sum_{r=\pm}\int_0^\infty dx
(\sqrt{x^2+w^2}+ru)^2\frac{e^{\sqrt{x^2+w^2}+ru}}{(e^{\sqrt{x^2+w^2}+ru}+1)^2},\nonumber\\
G(u,w)&=&\sum_{r=\pm}\int_0^\infty dx
\left[1+\frac{ru}{\sqrt{x^2+w^2}}\right]\frac{e^{\sqrt{x^2+w^2}+ru}}{(e^{\sqrt{x^2+w^2}+ru}+1)^2}.
\label{cv2}
\end{eqnarray}
Note that the contribution from the excitation branch with $a=+$ is
suppressed by a factor $e^{-\mu/T}$ at low temperature, it is
negligible except near the critical temperature $T_c$. At $T_c$ with
$w=0$, we recover $K(u,0)\simeq\pi^2/3$ in the normal phase. The
first term of $G(u,0)$ is approximately unit and leads to the
standard jump of specific heat at the critical temperature for a
symmetric superfluid without Fermi surface mismatch between the two
species. The second term of $G(u,0)$ reflects the gapless nature of
the asymmetric pairing. If one turns off the interaction
between quarks ($\kappa=0$), the $K$-term reproduces the low
temperature specific heat of two-flavor quark matter $c_{V0}=\gamma
T$ with $\gamma=E_u^Fp_u^F+E_d^Fp_d^F$ with the $u$- and $d$-quark
Fermi energies and Fermi momenta, while the $G$-term gives the jump of
the specific heat at the normal-superfluid transition point. The
numerical result of $c_V(T)$ is shown in Fig.\ \ref{cvT}. The quark
mass dependence of $c_V$ is weak and can be neglected.

\begin{figure}[!htb]
\includegraphics[width=7cm]{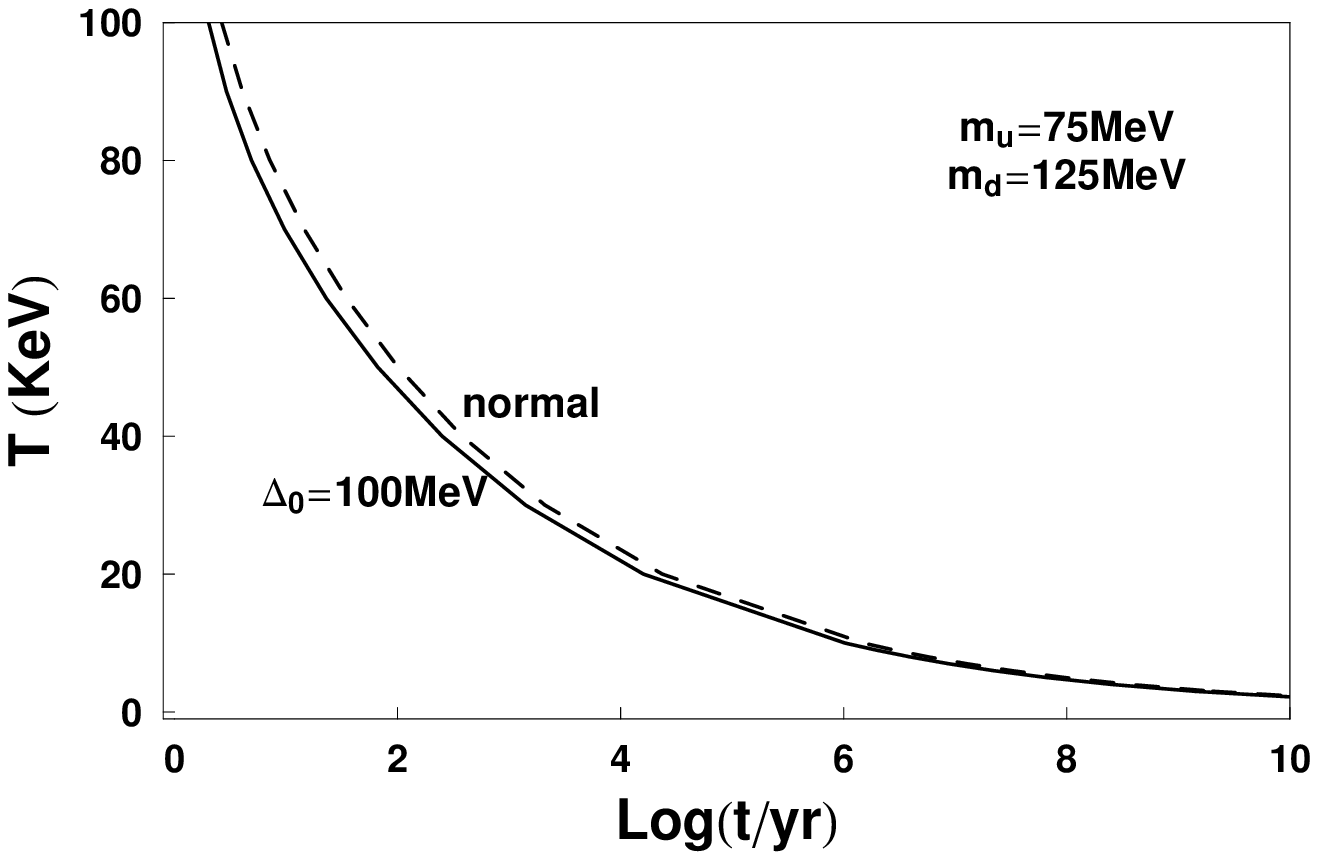}
\includegraphics[width=7cm]{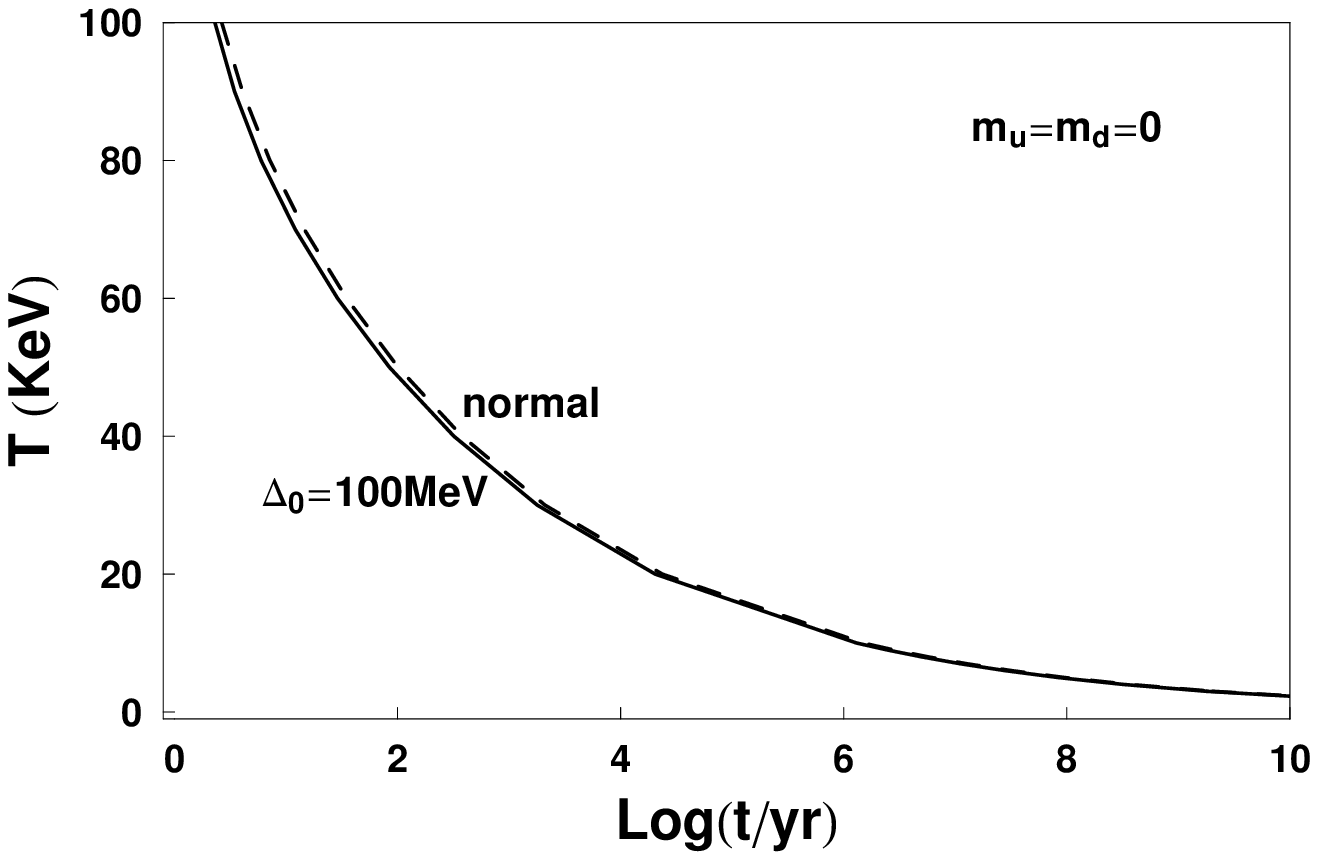}
\caption{The cooling curves in pion condensed (solid lines) and
normal (dashed lines) quark matter for massive (left panel) and
massless (right panel) quarks. }
\label{Tt}
\end{figure}

Substituting the neutrino emissivity (\ref{e4}) and the specific heat
(\ref{cv1}) into Eq. (\ref{rate}), we obtain
\begin{equation}
\label{rate2}
t-t_0=-\frac{\pi^3}{2G^2\mu_e}\int_{T_0}^T \frac{d\, T'}{T'^5}
\frac{p_u^0 E_u^0 + p_d^0 E_d^0}{E_d^0(E_u^0-p_u^0\cos{\theta_{ue}})}
\frac{K(u,w)+\Delta_0^2T_c^{-2}G(u,w)}{F(u,w)},
\end{equation}
where $p_{u,d}^0$ and $E_{u,d}^0$ depend on $T$ through $\Delta(T)$.
Fig.\ \ref{Tt} shows the time evolution of the pion condensed quark
matter with initial temperature $T_0=1$ MeV at $t_0=1$ yr. For
comparison we also show the cooling evolution in normal quark
matter. In the case with mass difference between $u$- and
$d$-quarks, the cooling of the condensed phase is faster.
This is due to the larger neutrino emissivity and smaller
specific heat in the pion condensed quark matter. For massless
quarks, the two cooling curves for condensed and normal
quark matter are almost the same due to the fact that the emissivity
difference between the two cases is small at low temperature, see
Fig.\ \ref{e-T}.

\section{Contributions from negative energy excitations}
\label{n-em}
As we have argued above that the positive energy excitations
dominate the neutrino emissivity at $\mu_B \gg \Delta$ and at low
temperature $T\ll \mu_{u,d}$. In principle, the contribution from
negative energy excitations must be considered if we follow the
exact calculation (\ref{e2ab}) at high temperature. In order to
better understand the role of negative energy excitations, we
discuss the emissivity at $T\rightarrow T_c$. In this case, the
$j,k,b,c$ related part of the integrands in Eq.\ (\ref{e2ab}) reads
\begin{eqnarray}
I_{\nu} &=& B_{r}^{a}(p_u) B_{-s}^{-b}(p_d) n_F(E_\nu + rE_u^a - sE_d^{-b})
n_F(-rE_{u}^{a}-\mu) n_F(sE_{d}^{-b}+\mu) |M_{ab}|^2 \nonumber\\
&\simeq& \Theta [-ra\mathrm{Sgn}(E_u+a\delta\mu )]
\Theta [-sb\mathrm{Sgn}(E_d-b\delta\mu )]
n_F(E_\nu + r|E_u+a\delta\mu |- s |E_d-b\delta\mu |)\nonumber\\
&&\times n_F(-r |E_u+a\delta\mu | - \mu)
n_F(s |E_d-b\delta\mu | + \mu) |M_{ab}|^2\nonumber\\
&=& n_F(E_\nu - a E_u + b E_d - \mu_e ) n_F(a E_u -\mu _u)
n_F(-b E_d + \mu _d) |M_{ab}|^2 , \nonumber\\
I_{\bar{\nu}} &\simeq & n_F(E_\nu + a E_u - b E_d + \mu_e ) n_F(-a E_u +\mu _u)
n_F(b E_d - \mu _d) |M_{ab}|^2 ,
\end{eqnarray}
where we have used the approximation $B_a^r(p_u)\simeq\Theta
\left[-ra\mathrm{Sgn}(E_u+a\delta\mu )\right]$ at $T\rightarrow
T_c$ which means that only the terms with
$r\mathrm{Sgn}(E_u+a\delta\mu )=-a$ are relevant in the
calculation of neutrino emission, the step and sign functions are
defined as $\Theta(x)=1$ for $x>0$ and $0$ for $x<0$, and
$\mathrm{Sgn}(x)=1$ for $x>0$ and $-1$ for $x<0$.

The above expressions with $a=+$ and $b=+$ reproduces the
conventional result for normal quark matter,
\begin{eqnarray}
I_{\nu}& \sim & n_F(E_\nu - E_u + E_d - \mu_e ) n_F(E_u -\mu _u)
n_F(- E_d + \mu _d) |M_{++}|^2, \nonumber\\
I_{\bar{\nu}} & \sim & n_F(E_\nu + E_u - E_d + \mu_e ) n_F(- E_u
+\mu _u) n_F( E_d - \mu _d) |M_{++}|^2 .
\end{eqnarray}
At low temperature $T\ll \mu_{u,d}$, this contribution dominates
the neutrino emission, since the integrals in (\ref{e2ab}) have
sharp peaks at $E_u\sim\mu_u$ and $E_d\sim\mu_d$ and the
contribution from negative energy excitations is suppressed at
least by a factor of $e^{-\mu _{u,d}/T}$. However, at temperature
$T\sim \mu_{u,d}$, such a suppression disappears, and those
processes which are forbidden by energy conservations at low
temperature can happen. For example, for $ab=-+$ the processes
with energy conservations $E_\nu + E_u + E_d =\mu_e$ and $E_u +
E_d=E_\nu+\mu_e$ for neutrinos and ant-neutrinos are forbidden at
low temperature, but allowed at $T\sim \mu_{u,d}$ since the quark
energies could be as large as the chemical potentials.

\section {Summary}
\label{summary}
Quark matter may exist in the core of a compact star. Pion
condensed phase is one possible ground state of the quark matter
due to the non-zero isospin density as a result of the $\beta$
equilibrium between $u$ and $d$ quarks. For the neutrino emission,
the role of the pion condensation in quark matter is different
from that in nuclear matter. In nuclear matter, the presence of
pion degrees of freedom effectively opens a new channel for the
nucleon interaction and results in an enhancement of the neutrino
emissivity. In this paper we considered the gapless pion
excitations which can be realized in compact stars. We calculated
the neutrino emissivity and cooling rate through direct Urca
processes in pion condensed quark matter with two flavors. We
derived general expressions for neutrino and anti-neutrino
emissivity. At temperature much lower than quark chemical
potentials, we demonstrated that the positive energy excitation
dominates the emissivity if the baryon chemical potential is much
larger than the magnitude of the condensate. We also discussed the
role of negative energy excitations at temperature comparable to
the chemical potential.

We showed the numerical results for the neutrino emissivity and
cooling rate at low temperature. The total neutrino emissivity is
of the same order as that in the normal phase, which is a
result of gapless excitations. While the results in the two phases
are quite close to each other when quarks are massless, the mass
difference between $u$- and $d$-quarks due to isospin symmetry
breaking leads to a remarkable enhancement of the emissivity. For
typical values of the initial temperature, the chemical potential and the
pion condensate corresponding to the compact star interior, the
cooling of the condensed phase is found to be faster
than that of the normal one, when the quark mass
splitting is sizable.

\vspace{0.5cm}

{\bf Acknowledgments:} The work is supported by the NSFC Grants
No. 10575058, 10428510 and 10675109 and the startup grant from
University of Science and Technology of China in association with
the 'Bai Ren' project of Chinese Academy of Sciences.

\appendix
\section{Asymptotic behavior of $F(u,w)$ at low temperature}
\label{asymp} In the limit of $u > w \gg 1$ at low temperature,
the four components of $F(u,w)$ can be simplifies as
\begin{eqnarray}
F_{++}(u,w) & \simeq & \int_{0}^{\infty}dv dx dy v^3
\left[e^{\sqrt{y^{2}+w^{2}}+u}+e^{\sqrt{x^{2}+w^{2}}+u+v}\right]^{-1}\nonumber \\
& \simeq & e^{-u-w}\int_{0}^{\infty}dv dx dy v^3
\left[e^{y^{2}/(2w)}+e^{x^{2}/(2w)+v}\right]^{-1}\nonumber\\
& \simeq & we^{-u-w},\nonumber\\
F_{+-}(u,w) & \simeq & \int_{0}^{\infty} dv dx dy v^3
\left[e^{\sqrt{x^{2}+w^{2}}+\sqrt{y^{2}+w^{2}}+v}+e^{\sqrt{x^{2}+w^{2}}+u+v}\right]^{-1}\nonumber\\
& \simeq & \int_{0}^{\infty}dv dx dy v^3
e^{-\sqrt{x^{2}+w^{2}}-v}\left[e^{\sqrt{y^{2}+w^{2}}}+e^{u}\right]^{-1}\nonumber \\
& \simeq & e^{-w-u}\int_{0}^{\infty}dv dx dy v^3 e^{-x^{2}/(2w)-v}
\left[e^{\sqrt{y^{2}+w^{2}}-u}+1\right]^{-1}\nonumber \\
& \simeq & 3\sqrt{2\pi
w}e^{-w-u}\int_{0}^{\sqrt{u^{2}-w^{2}}}dy\nonumber\\
&\simeq& w^{1/2}(u^2-w^2)^{1/2} e^{-u-w},\nonumber\\
F_{-+}(u,w) & \simeq & \frac{1}{4}\int_{0}^{\infty}dx dy dv v^3
\left[ e^{\sqrt{x^{2}+w^{2}}-u}+1\right]^{-1}
\left[ e^{\sqrt{y^{2}+w^{2}}+u}+1\right]^{-1}
\left[\sqrt{x^{2}+w^{2}}+\sqrt{y^{2}+w^{2}}\right]^4\nonumber \\
& \simeq &
\frac{1}{4}\int_{0}^{\sqrt{u^{2}-w^{2}}}dx\int_{0}^{\infty}dy
e^{-\sqrt{y^{2}+w^{2}}-u}\left[\sqrt{x^{2}+w^{2}}+\sqrt{y^{2}+w^{2}}\right]^4\nonumber\\
& \simeq & w^{1/2}(u^2-w^2)^{9/2} e^{-u-w},\nonumber\\
F_{--}(u,w) & \simeq &
\int_{0}^{\sqrt{u^{2}-w^{2}}}dx\int_{\sqrt{u^{2}-w^{2}}}^{\infty}dy
\int_{0}^{\infty}dv\frac{v^{3}}{e^{-\sqrt{x^{2}+w^{2}}+\sqrt{y^{2}+w^{2}}+v}+1}\nonumber\\
& \simeq & O(1),
\end{eqnarray}
where we have used
\begin{equation}
\mathrm{Lim}_{T\rightarrow 0}\frac{1}{e^{x/T}+1} \simeq
\Theta(-x).
\end{equation}
The dominant contribution to the integral over $x$ in $F_{--}$
comes from the regions near the upper limit
$x=\sqrt{u^{2}-w^{2}}-0^{+}$ and lower limit
$x=\sqrt{u^{2}-w^{2}}+0^{+}$.

\section{exact calculation of emissivity for neutrinos and anti-neutrinos}
\label{exact} 
\begin{figure}[!htb]
\includegraphics[width=7cm]{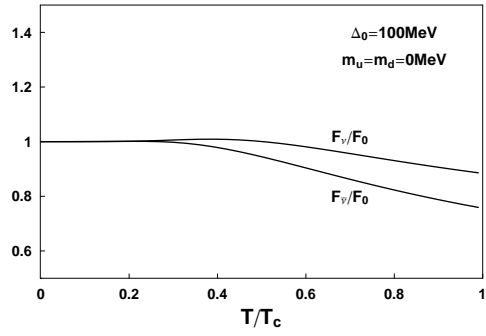}
\caption{The scaled $F_\nu$ and $F_{\bar\nu}$ as functions of
scaled temperature $T/T_c$. }\label{fnu}
\end{figure}

In this appendix, we calculate exactly the neutrino
and anti-neutrino emissivity listed in (\ref{e3ab}), in order to see
how good is the approximation we made in deriving (\ref{e4}). From
(\ref{e3ab}), the pre-factors in $\epsilon_\nu$ and
$\epsilon_{\bar\nu}$ are exactly the same, and the difference is
from the summations and integrals. We denote the summation and
integration in $\epsilon_\nu$ by $F_\nu$ and that in
$\epsilon_{\bar\nu}$ by $F_{\bar\nu}$. It is easy to see the
relation $F_\nu(-\delta\mu)=F_{\bar\nu}(\delta\mu)$. In fact, it is
true even for the original $\epsilon_\nu$ and $\epsilon_{\bar\nu}$
in (\ref{e2ab}). For $\mu\gg\delta\mu$, the two integrals are
dominated by the phase space near gapless momenta at low
temperature, the $\delta\mu$ dependence of $F_\nu$ and
$F_{\bar{\nu}}$ is negligible, and we have therefore
$F_\nu=F_{\bar{\nu}}$. However, if the temperature is not very low
or $\delta\mu$ approaches to $\mu$, the gapless regions might not
dominate the integrals, and the explicit dependence on $\delta\mu$
must enter $F_\nu$ and $F_{\bar{\nu}}$. In Fig.\ \ref{fnu}, the
exact $F_\nu$ and $F_{\bar\nu}$ scaled by $F_0\equiv
F_\nu(\Delta_0=0,T=0)$ are presented as functions of $T/T_c$. While
at low temperature $F_\nu$ and $F_{\bar\nu}$ coincide very well,
indicating that the approximation we used is good, the difference
between $F_\nu$ and $F_{\bar\nu}$ increases as the temperature grows
and reaches $10\%$ at $T\rightarrow T_c$.

\section{parameter expansion near $T_c$}
\label{app1} At $T \rightarrow T_c$ in the pion condensation, to the
leading order of $\kappa, m^2/\mu^2, \Delta ^2/\mu^2$, the gapless
energies and momenta of quarks and the corresponding angle between
$u$-quark and electron momenta can be written as
\begin{eqnarray}
E_u^0 &\approx & \mu_u\left(1-\frac{\Delta^2}{2\mu\mu_u}\right),\nonumber\\
E_d^0 &\approx & \mu_d\left(1-\frac{\Delta^2}{2\mu\mu_d}\right),\nonumber\\
p_u^0 &\approx &\mu_u \left[ 1-\kappa-\frac{m_u^2}{2\mu_u^2}
-\frac{\Delta^2}{2\mu_u\mu}\right],\nonumber\\
p_d^0 &\approx &\mu_d \left[ 1-\kappa-\frac{m_u^2}{2\mu_d^2}
-\frac{\Delta^2}{2\mu_d\mu}\right],\nonumber\\
\cos{\theta_{ue}^0} &\approx &
1-\kappa\frac{\mu_d}{\mu_u}+\frac{m_u^2}{2\mu_u^2}-
\frac{m_d^2-m_u^2}{2\mu_u\mu_e} ,
\end{eqnarray}
which lead to
\begin{eqnarray}
E_u^0 E_d^0\left(1-\frac{p_u^0}{E_u^0}\cos{\theta_{ue}^0} \right)
&\approx &\mu_u\mu_d\left(1-\frac{\Delta^2}{2\mu\mu_u}\right)
\left(1-\frac{\Delta^2}{2\mu\mu_d}\right)\nonumber\\
&&\times\left[1-\left(1-\kappa-\frac{m_u^2}{2\mu_u^2} \right)
\left(1-\kappa\frac{\mu_d}{\mu_u}+\frac{m_u^2}{2\mu_u^2}-
\frac{m_d^2-m_u^2}{2\mu_u\mu_e} \right) \right]\nonumber\\
&\approx & \mu_u\mu_d\left(1-\frac{\Delta^2}{\mu_u\mu_d}\right)
\left[\left(1+\frac{\mu_d}{\mu_u}\right)\kappa
+\frac{m_d^2-m_u^2}{2\mu_u\mu_e}\right].
\end{eqnarray}
Substituting this relation into (\ref{e4}), we obtain (\ref{e5}).

\end{document}